\documentclass[11pt,a4paper]{article}

\usepackage{jcappub}
\usepackage{amssymb,amsmath,graphicx,latexsym}
\usepackage{bm}
\usepackage{epsfig}

\def\beq{\begin{equation}}
\def\eeq{\end{equation}}
\def\ber{\begin{eqnarray}}
\def\eer{\end{eqnarray}}
\def\benu{\begin{enumerate}}
\def\eenu{\end{enumerate}}
\def\nn{\nonumber}
\def\l{\left}
\def\r{\right}
\def\d{{\rm d}}

\def\pa{\partial}
\def\f{\frac}
\def\mpl{M_{p}}

\def \lleq {\lower0.9ex\hbox{ $\buildrel < \over \sim$} ~}
\def \ggeq {\lower0.9ex\hbox{ $\buildrel > \over \sim$} ~}

\def\prd{{Phys.\@ Rev.\@ D\ }}

\title{Resurrecting power law inflation in the light of \emph{Planck} results}
\author[a]{Sanil Unnikrishnan}
\author[b]{and Varun Sahni}

\affiliation[a]{School of Physics, Indian Institute of Science Education and Research,
Thiruvananthapuram 695016, India}
\affiliation[b]{Inter-University Centre for Astronomy and Astrophysics,
Post Bag 4, Ganeshkhind, Pune 411~007, India}

\emailAdd{sanil@iisertvm.ac.in}
\emailAdd{varun@iucaa.ernet.in}

\date{\today}

\abstract{
It is well known that a canonical scalar field with an exponential potential can drive power law inflation (PLI).
However, the tensor-to-scalar ratio in such models turns out to be larger than the stringent limit set by recent \emph{Planck} results. We propose a new model of power law inflation for which the scalar spectra index, the tensor-to-scalar ratio and the non-gaussianity parameter $f_{_{\mathbf{NL}}}^{\mathrm{equil}}$ are in excellent agreement with \emph{Planck} results. Inflation, in this model, is driven by a  non-canonical scalar field with an {\em inverse power law} potential. The Lagrangian for our model is structurally similar to that of a canonical scalar field and has a power law form for the kinetic term. A simple extension of our model resolves the graceful exit problem which usually afflicts models of power law inflation.}

\arxivnumber{1305.5260}

\begin{document}
\maketitle

\section{Introduction}\label{sec:intro}

It is widely believed that the early universe underwent a brief `inflationary phase' during which
its expansion rapidly accelerated. The inflationary paradigm was originally suggested to resolve the horizon, flatness
and singularity problems of Big Bang cosmology~\cite{models}. An attractive byproduct of the paradigm was its ability to provide an elegant causal mechanism to seed the formation of large scale structure in the universe~\cite{pert}.
An important early prediction of inflation -- that of a nearly scale invariant spectrum of perturbations -- has been spectacularly confirmed by experiments of the Cosmic Microwave Background (CMB) including the current \emph{Planck} mission~\cite{Planck-inflation,Planck-cosmo-parameters}. Indeed, increasingly accurate measurements of the CMB have significantly reduced the number of theoretical models in the inflationary zoo~\cite{Martin-2013,Riotto-2002,Martin-2004,Bassett-2005,Kinney-2009,Sriram-2009,Baumann-2009}, and one hopes that current and future CMB experiments will further help in pointing the way forward for inflationary model building~\cite{wang-2013}.

An inflationary model is usually characterized by observables including:
(1) the scalar spectral index $n_{_{S}}$, (2) tensor-to-scalar ratio $r$~\cite{Starobinsky-1979,Starobinsky-1985},
(3) the non-gaussianity parameter $f_{_{\mathbf{NL}}}$. Other important inflationary parameters include the running of scalar spectral index $n_{_{\mathrm{run}}} \equiv  {\rm d} n_{_{S}}/{\rm d~ln~}k$ and the spectral index for tensor perturbations $n_{_{T}}$. Recent CMB data from \emph{Planck} combined with the large angle polarization data from the Wilkinson Microwave Anisotropy Probe (WMAP) (henceforth {\em Planck} + WP) place strong bounds on these parameters:
$n_{_{S}}\,=\,0.9603\,\pm\,0.0073$ and $r\,<\,0.11$ at $95\%$ CL~\cite{Planck-inflation}. Since \emph{Planck} data does not indicate any statistically significant running of the spectral index~\cite{Planck-inflation}, it is interesting to investigate the viability of power law inflation for which $n_{\mathrm{run}}$ {\em identically vanishes}. Although, there do exist other models with constant $n_{_{S}}$ (see for instance, Ref.~\cite{Starobinsky-2005} for an exact non-power law model of inflation  which gives $n_{_{S}}\,=\,1$ and Ref.~\cite{Vallinotto-2004} for models which give approximately constant $n_{_{S}}$), our analysis will be confined to the simplest case, \emph{viz}.~power law inflation.

Power law inflation (PLI) with $a(t) \propto t^q, ~q > 1$ arises in the context of a canonical scalar field with an exponential potential~\cite{Abbott-1984,Lucchin-1985,sahni88}. The  tensor-to-scalar ratio in canonical PLI, originally determined in \cite{ss92}, turns out to be larger than the limits set by \emph{Planck} data,
which rules out this class of models.
PLI can also arise in a K-inflation scenario~\cite{Garriga-1999} with the Lagrangian of the form $ {\cal L}(\phi , X) = g(X)f(\phi)$  where $X = (1/2)\pa_{\mu}\phi\; \pa^{\mu}\phi$. A simple quadratic function for $g(X)$ such as the one proposed by  Armendariz-Picon {\it et.~al.~}\cite{Picon-1999} \emph{viz.\ }  $g(X) = -X\,+ X^{2}$ gives acceptable $n_{_{S}}$ and $r$ but comes into tension with \emph{Planck} data~\cite{Planck-NG} on account of the large value of the non-gaussianity parameter $f_{_{\mathbf{NL}}}^{\mathrm{equil}}$, determined in the equilateral limit~\cite{chen-2007}.
However a suitable choice of $g(X)$ can alleviate this problem by altering the speed of sound
so that the non-gaussianity parameter $f_{_{\mathbf{NL}}}^{\mathrm{equil}}$ falls within
an acceptable range.

In this paper we propose a new PLI model in which inflation is driven by a non-canonical scalar field  with the Lagrangian~\cite{our-JCAP-paper} $ {\cal L}(\phi , X) = X^\alpha - V(\phi)$. In this model observables such as $n_{_{S}}$, $r$ and $f_{_{\mathbf{NL}}}^{\mathrm{equil}}$ lie well within the limits set by the \emph{Planck} data, for a wide range of model parameters.

This paper is organized as follows. Sec.~\ref{sec:Standard PLI} briefly reviews power law inflation,
with Sec.~\ref{sec:canonical PLI} focussing on PLI within the canonical framework.
Our new model, based on a non-canonical Lagrangian with an {\em inverse power law}
potential, is described  in Sec.~\ref{sec:new PLI}. It is well known that PLI models suffer from a {\em graceful exit problem} since inflation continues forever in such models precluding the possibility of reheating. Our new model does not share this drawback since it can accommodate a graceful exit, as demonstrated in Sec.~\ref{sec:exit}. Our main results are summarized in  Sec.~\ref{sec:conclusions}.

\section{Power law inflation}
\label{sec:Standard PLI}
In a spatially flat Friedmann-Robertson-Walker (FRW) universe
\beq
\d s^2 = \d t^2-a^{2}(t)\,\l[\d x^2 + \d y^2 + \d z^2\r],
\label{eqn: FRW}
\eeq
power law expansion corresponds to
\beq
a(t)\, \propto\, t^{q}~,
\label{eqn: a(t)}
\eeq
where $q\,>\,1$ for power law inflation (PLI).
The slow roll parameters $\varepsilon,\delta$ are defined as
\beq
\varepsilon \equiv -\frac{\dot{H}}{\,H^{2}}~,~~\delta \equiv \varepsilon -  \frac{\dot{\varepsilon}}{2\,H\,\varepsilon}~,
\label{eqn: slow roll p 1}
\eeq
where $H \equiv \dot{a}/a$.
It is easy to see that for power law expansion $H = q/t$ so that
\beq
\delta = \varepsilon = q^{-1}~.
\eeq
Slow roll PLI corresponds to $\varepsilon << 1$ which occurs when $q >> 1$.

The fact that $\varepsilon$ is identically constant could be viewed as a drawback of PLI since it makes exit from inflationary expansion difficult. PLI models are therefore somewhat incomplete since an exit mechanism needs to be added  in order for a decelerating phase to succeed inflation.  A resolution of this quandary, in the form of a reasonable exit mechanism, will  be  discussed later in Sec.~\ref{sec:exit}. For the moment we shall assume that such an exit mechanism will not significantly alter the power law nature of the solution (\ref{eqn: a(t)}) during inflation.

This paper shall focus on the action
\beq
S[\phi]=\int\!\d^{4}x\, \sqrt{-g}\; {\cal L}(X,\phi),\label{eqn: action}
\eeq
where the Lagrangian density ${\cal L}(\phi , X)$ can, in general,  be an arbitrary function of the field $\phi$ and the kinetic term $X~=~(1/2)\pa_{\mu}\phi\; \pa^{\mu}\phi$.
For a generic ${\cal L}(\phi , X)$, besides the two slow roll parameters (\ref{eqn: slow roll p 1}) it is convenient to introduce a third slow roll parameter $\sigma$ defined as~\cite{Hu-2011}
\beq
\sigma = \frac{\dot{c_{_s}}}{H\,c_{_s}}~,\label{eqn: sigma}
\eeq
where $c_{_s}$ is the speed of sound of the scalar field~\cite{Garriga-1999}
\beq
c_{_s}^{2} \equiv \l[\f{\l({\pa {\cal L}}/{\pa X}\r)}{\l({\pa {\cal L}}/{\pa X}\r)
\,+\, \l(2\, X\r)\l({\pa^{2} {\cal L}}/{\pa X^{2}}\r)}\r].
\label{eq:sound_speed_def}
\eeq

Slow roll inflation requires not only $\varepsilon <<1$ and $|\delta| << 1$ but also $|\sigma| << 1$. For the canonical scalar field the value of $\sigma$ is identically zero and this is also the case for kinetically driven PLI~\cite{Picon-1999} as well as the non-canonical model~\cite{our-JCAP-paper} studied in this paper. Some of the results of this paper will remain valid for any PLI model based on the action~(\ref{eqn: action}) but for which $\sigma$ identically vanishes. The PLI scenario for the case when $\sigma$ is non-zero but  $|\sigma| << 1$ will also be briefly mentioned in this paper.

Broadly speaking, the functional form of the Lagrangian density ${\cal L}(\phi , X)$ can be divided into two types: (1) ${\cal L}(\phi , X) = F(X) - V(\phi)$~\cite{De-Santiago-2013} and (2) ${\cal L}(\phi , X) = F(X)V(\phi)$~\cite{Garriga-1999}, although other possibility also exist, for instance DBI models~\cite{DBI}. Canonical scalar field models are the simplest possible formulation of the first type with ${\cal L}(\phi , X) = X - V(\phi)$, while the model proposed by Garriga and Mukhanov~\cite{Garriga-1999}, and by  Armendariz-Picon {\it et.~al.~}\cite{Picon-1999} is of the second type. Our non-canonical model~\cite{our-JCAP-paper} is of the first type with $F(X) \propto X^{\alpha}$.

\subsection{Power law inflation from a canonical scalar field}
\label{sec:canonical PLI}

It is well known that a canonical scalar field with the Lagrangian density~\cite{Lucchin-1985}
\beq
{\cal L}(X,\phi) = \frac{1}{2}\pa_{\mu}\phi\,\pa^{\mu}\phi\; -\; V_{_0}\exp\l[-\sqrt{\frac{2}{q}}\,\l(\frac{\phi}{\mpl}\r)\r],
\label{eqn: can-Lagrangian}
\eeq
can drive power law expansion with $a(t) \propto t^q$, in a spatially flat FRW space-time ($\mpl \equiv1/\sqrt{8\pi G}$).

The power spectrum of scalar curvature perturbation $\mathcal{R}_{_k}$ is defined as
\beq
\mathcal{P}_{_{S}}(k) \equiv \l(\frac{k^{3}}{2\pi^{2}}\r)|\mathcal{R}_{_k}|^{2}~,
\eeq
while the tensor power spectrum is
\beq
\mathcal{P}_{_{T}}(k) \equiv 2\l(\frac{k^{3}}{2\pi^{2}}\r)|h_{_k}|^{2},
\label{eq:tensor}
\eeq
where $h$ is the amplitude of the tensor perturbation. The scalar spectral index $n_{_{S}}$  and the corresponding spectral index for the tensor perturbations are defined as
\ber
n_{_{S}} - 1 &\equiv&  \frac{\d\, \mathrm{ln} \mathcal{P}_{_{S}}}{\d\, \mathrm{ln} k},
\label{eqn: ns definition}\\
n_{_{T}} &\equiv&  \frac{\d\, \mathrm{ln} \mathcal{P}_{_{T}}}{\d\, \mathrm{ln} k}~.
\label{eqn: nt definition}
\eer
Furthermore, the tensor-to-scalar ratio is defined as
\beq
r \equiv \f{\mathcal{P}_{_{T}}}{\mathcal{P}_{_{S}}}.
\label{eqn: T-to-S def}
\eeq

For the canonical  PLI model~(\ref{eqn: can-Lagrangian}), in the slow roll limit, one finds
\ber
n_{_{T}}\,=\, n_{_{S}} - 1\, &\simeq&\,  -\frac{2}{q}
\label{eqn: ns-can}\\
r\, &\simeq&\, \frac{16}{q}\,.\label{eqn: r-can}
\eer
Note that when slow roll is not assumed one gets $n_{_{S}}\,=\,1\,-\,2/(q-1)$. \emph{Planck} data in combination with the large angle polarization data from WMAP requires the value of the scalar spectral index $n_{_{S}}$ to lie in the range $[0.945-0.98]$ at $95\%$ CL~\cite{Planck-inflation}. This restricts $q$ in the power law solution $a(t)\, \propto\, t^{q}$ to the range
\beq
38\,\lesssim\,q\,\lesssim\,101\,.\label{eqn: q-range}
\eeq
With $q$ within the above range, the tensor-to-scalar ratio $r = 16/q$ lies in the range $0.16\,<\,r\,<0.43$. This is well above the limit set by Planck data which indicate $r\,<\,0.12$ at $95\%$ CL when BAO data is also included~\cite{Planck-inflation}. Therefore one concludes that PLI based on a canonical scalar field with an exponential potential is in considerable tension with \emph{Planck} data.

\section{A new model of power law inflation }
\label{sec:new PLI}
Consider a non-canonical scalar field with Lagrangian density~\cite{our-JCAP-paper,Li-2012}:
\beq
{\cal L}(X,\phi) = X\l(\frac{X}{M^{4}}\r)^{\alpha-1} -\; V(\phi),
\label{eqn: Lagrangian}
\eeq
where $\alpha$ is  dimensionless  while $M$ has dimensions of mass. For $\alpha = 1$, Eq.~(\ref{eqn: Lagrangian}) reduces to the usual canonical scalar field Lagrangian ${\cal L}(X,\phi) = X - V(\phi)$. Therefore (\ref{eqn: Lagrangian}) can be viewed as a natural generalization of the standard canonical scalar field Lagrangian.

It is interesting to note that, in contrast to the canonical case, chaotic inflation with $\alpha > 1$ and $V(\phi) = \lambda\phi^{4}$ in (\ref{eqn: Lagrangian}), agrees quite well with CMB data even for $\lambda$ as large as unity~\cite{our-JCAP-paper}; also see Ref.~\cite{Li-2012}.

We now reconstruct the potential $V(\phi)$ which can drive power law inflation for the Lagrangian density (\ref{eqn: Lagrangian}). The energy density, $\rho_{_{\phi}}$, and pressure, $p_{_{\phi}}$, are given by
\ber
\rho_{_{\phi}} &=& \l(\f{\pa {\cal L}}{\pa X}\r)\, (2\, X)- {\cal L}\label{eqn: rho-phi},\\
p_{_{\phi}} &=& {\cal L}\label{eqn: p-phi},  ~~
X \equiv \frac{1}{2} {\dot \phi}^{2}~.
\eer
Substituting for ${\cal L}$ from (\ref{eqn: Lagrangian}) results in
\ber
\rho_{_{\phi}} &=& \l(2\alpha-1\r)X\l(\frac{X}{M^{4}}\r)^{\alpha-1} +\;  V(\phi),\nonumber\\
p_{_{\phi}} &=& X\l(\frac{X}{M^{4}}\r)^{\alpha-1} -\; V(\phi)~.
\label{eqn: rho-p-model}
\eer
For the power law solution $a(t)\, \propto\, t^{q}$, the Friedmann equations
\ber
\l(\frac{\dot{a}}{a}\r)^{2} &=& \l(\frac{8 \pi G}{3}\r)\rho_{_{\phi}},\label{eqn: Friedmann eqn1}\\
\frac{\ddot{a}}{a} &=& -\l(\frac{4 \pi G}{3}\r)\l(\rho_{_{\phi}} + 3\,p_{_{\phi}}\r)\,,\label{eqn: Friedmann eqn2}
\eer
imply
\ber
\rho_{_{\phi}} &=& \frac{3\,\mpl^2\,q^{2}}{t^{2}}~,\\
p_{_{\phi}} &=& w_{_{\phi}}\,\rho_{_{\phi}}~, ~~
\label{eqn:rho-t}
\eer
where the equation of state parameter $w_{_{\phi}}$ is related to $q$ as follows
\beq
q = \frac{2}{3\,(1 + w_{_{\phi}})}~.
\label{eqn: q-w}
\eeq
From Eqs.~(\ref{eqn: rho-p-model}) to (\ref{eqn: q-w}) we get
\ber
\frac{d}{dt}\l(\frac{\phi}{\mpl}\r) &=& \mpl\,\l(\frac{q\,2^{\alpha}\,\mu^{4(\alpha-1)}}{\alpha}\r)^{\frac{1}{2\,\alpha}}\,\l[\mpl\,t\r]^{-\frac{1}{\alpha}},
\label{eqn: dphi-dt}\\
V(\phi)  &=& 3\,\mpl^{4}\,q^{2}\l(\frac{1 - (2\alpha-1)w_{_\phi}}{2\,\alpha}\r)\l[\mpl\,t\r]^{-2},~~~~
\label{eqn: V-t}
\eer
where
\beq
\mu = \frac{M}{\mpl}.
\label{eqn: mu}
\eeq
For a canonical scalar field ($\alpha =1$) integrating (\ref{eqn: dphi-dt}) gives $t(\phi) \propto \exp\l[\phi/(\mpl\sqrt{2q}\r]$. Substitution in (\ref{eqn: V-t}), results in the usual exponential potential for canonical PLI \emph{viz.~}$V(\phi) = V_{_0}\exp\l[-\sqrt{2/q}(\phi/\mpl)\r]$. Interestingly, when $\alpha \neq 1$, Eqs.~(\ref{eqn: dphi-dt}) and (\ref{eqn: V-t}) imply the {\em inverse power law} potential\footnote{In the canonical context ($\alpha=1$), the potential $V \propto \phi^{-s}$ gives rise to `intermediate' inflation~\cite{barrow90,muslimov90} with $a \propto \exp{(A\,t^p)}$ where $p = 4/(4+s)$. Such models are in tension with the {\em Planck}+WP+BAO results~\cite{Planck-inflation}.}~\cite{sanil-2008}
\beq
V(\phi)  = \frac{V_{_0}}{(\phi/\mpl)^{s}}~;~~~~\mathrm{where}~~~s  = \frac{2\,\alpha}{\alpha - 1}~.
\label{eqn: V-phi}
\eeq
In the above equation the constant $V_{_0}$ is given by
\beq
V_{_0} = M^{4}\l(\frac{1 - (2\alpha-1)w_{_\phi}}{2\,\alpha}\r)\l(\frac{s^{2\alpha}\,(3\,q)^{2\alpha - 1}}{\alpha\,6^{\alpha}}\r)^{\frac{1}{\alpha - 1}}~.
\label{eqn: V-0}
\eeq
Note that since $w_{_\phi}$ and $q$ are related via (\ref{eqn: q-w}), while $\alpha$ and $s$ are related via
(\ref{eqn: V-phi}), $V_{_0}$ is a function of three independent quantities, namely $M, \alpha$ and $q$. We therefore conclude that an inverse power law potential of the form (\ref{eqn: V-phi}) leads to power law expansion $a(t)\, \propto\, t^{q}$. It is important to note that since the parameter  $s$ in (\ref{eqn: V-phi}) does not dependent of the value of $q$, therefore it is the amplitude of the potential, $V_{_0}$, and not its shape, which determines the value of $q$ in $a(t)\, \propto\, t^{q}$.

In the slow roll limit, $\varepsilon \ll 1$, $|\delta| \ll 1$, the slow roll parameters~(\ref{eqn: slow roll p 1}) can be related to the  potential $V$ as follows~\cite{our-JCAP-paper}:
\beq
\varepsilon \simeq \varepsilon_{_V} = \l[\frac{1}{\alpha}
\l(\frac{3\,M^{4}}{V}\r)^{\alpha -1}
\l(\frac{\mpl\,V'}{\sqrt{2}\;V}\r)^{2\alpha}\r]^{\frac{1}{2\alpha - 1}}~,
\label{eqn: potential SR1}
\eeq
where the subscript in $\varepsilon_{_V}$ indicates that this parameter depends only on the potential. For the inverse power law potential (\ref{eqn: V-phi}), one finds $\varepsilon_{_V} \simeq 1/q$. Note that in the canonical case $\alpha = 1$, and $\varepsilon_{_V}$ reduces to its well known form
\beq
\varepsilon_{_V}^{(c)} = \frac{\mpl^{2}}{2}\l(\frac{V'}{V}\r)^{2}~.\label{eq:standard_slowroll}
\eeq

During slow roll, the power spectrum of scalar curvature perturbations in our model~(\ref{eqn: Lagrangian}) acquires the form~\cite{our-JCAP-paper}
\beq
\mathcal{P}_{_{S}}(k)\, =\, \l(\frac{1}{72\pi^{2}c_{_s}}\r)
\l\{\l(\frac{\alpha\, 6^{\alpha}}{\mu^{4(\alpha-1)}}\r)\l(\frac{1}{\mpl^{14\alpha -8}}\r)\l(\frac{V(\phi)^{5\alpha - 2}}{V'(\phi)^{2\alpha}}\r)
\r\}^{\frac{1}{2\alpha - 1}}_{a\,H\, =\, c_{_S}k}~,
\label{eqn: scalar PS model}
\eeq
where $c_{_s}$ is the speed of sound defined in (\ref{eq:sound_speed_def}), which, for our model (\ref{eqn: Lagrangian}) turns out to be
\beq
c_{_S} = \f{1}{\sqrt{2\,\alpha - 1}}.
\label{eqn: sound speed model}
\eeq
In Eq. (\ref{eqn: scalar PS model}), the quantity on the right hand side must be evaluated at the sound horizon exit $a\,H = c_{_S}k$. Substituting the expression for $V(\phi)$ from Eq.~(\ref{eqn: V-phi}) into Eq.~(\ref{eqn: scalar PS model}) and using Eq.~(\ref{eqn: V-0}), we get
\beq
\mathcal{P}_{_{S}}(k) = \l(\frac{q}{24\pi^{2}c_{_s}}\r)\l(\frac{1 - (2\alpha-1)w_{_\phi}}{2\,\alpha}\r)^{\frac{\alpha-1}{2\alpha-1}}
\l(\frac{V(\phi)}{\mpl^{4}}\r)_{a\,H\, =\, c_{_S}k}\label{eqn: scalar PS1}
\eeq
Since $a(t)\, \propto\, t^{q}$, we can write $a(t) = a_{i}\l(t/t_{i}\r)^{q}$. Therefore, at sound horizon exit $a\,H = c_{_S}k$ one gets
\beq
\mpl\,t =  \l(\mpl\,t_{i}\r)\l(\frac{t_{i}\,k_{\ast}}{a_{i}\,q}\r)^{\frac{1}{q - 1}}\l[c_{_S}\r]^{\frac{1}{q - 1}}\l(\frac{k}{k_{\ast}}\r)^{\frac{1}{q - 1}},
\label{eqn: mpl-t-k}
\eeq
where $k_{\ast}$ is an arbitrary wavenumber which can be set as the pivot scale. Substituting the above expression in Eq.~(\ref{eqn: V-t}) we get
\beq
\l(\frac{V(\phi)}{\mpl^{4}}\r)_{a\,H\, =\, c_{_S}k}\,=\, g\,\l[c_{_S}\r]^{-\frac{2}{q - 1}}\l(\frac{k}{k_{\ast}}\r)^{-\frac{2}{q - 1}},\label{eqn: V-at-SH-exit}
\eeq
where $g$ is defined as
\beq
g\,=\, 3\,q^{2}\l(\frac{1 - (2\alpha-1)w_{_\phi}}{2\,\alpha}\r)\l[\l(\mpl\,t_{i}\r)^{q-1}\l(\frac{t_{i}k_{\ast}}{a_{i}\,q}\r)\r]^{-\frac{2}{q-1}}.
\label{eqn: g}
\eeq
From Eqs.~(\ref{eqn: scalar PS1}) and (\ref{eqn: V-at-SH-exit}), the expression for the scalar power spectrum turns out to be
\beq
\mathcal{P}_{_{S}}(k)\, =\,A_{_S} \l(\frac{k}{k_{\ast}}\r)^{-\frac{2}{q - 1}},
\label{eqn: scalar PS}
\eeq
where
\beq
A_{_S} = \l(\frac{q}{24\pi^{2}c_{_s}}\r)\l(\frac{1 - (2\alpha-1)w_{_\phi}}{2\,\alpha}\r)^{\frac{\alpha-1}{2\alpha-1}}g\,\l[c_{_S}\r]^{-\frac{2}{q - 1}}~.
\label{eqn: AS}
\eeq
In the slow roll limit ($q\, >> \, 1$), the above expression simplifies to
\beq
A_{_S} \simeq \l(\frac{q\,g}{24\pi^{2}c_{_s}}\r)~.
\label{eqn: AS-SR limit}
\eeq

The expression for the tensor perturbation in terms of the potential $V(\phi)$ in our model~(\ref{eqn: Lagrangian}) is exactly the same as the one for the canonical scalar field in the slow roll limit and is given by~\cite{Garriga-1999}
\beq
\mathcal{P}_{_{T}}(k) = \l(\frac{2}{3\,\pi^{2}}\r)\l(\frac{V(\phi)}{\mpl^{4}}\r)_{aH\, =\, k}~.
\label{eqn: Tensor PS model}
\eeq
It is important to note that unlike the expression (\ref{eqn: scalar PS model}) which one evaluates at the sound horizon exit $a\,H = c_{_S}k$, the  tensor power spectrum expression (\ref{eqn: Tensor PS model}) must be evaluated at the horizon exit $a\,H = k$~\cite{Garriga-1999}.
Since the expression (\ref{eqn: V-at-SH-exit}) is valid for any value of $c_{_S}$, we can evaluate the quantity  $(V(\phi)/\mpl)$ at horizon exit by setting $c_{_S} =  1$ in Eq.~(\ref{eqn: V-at-SH-exit}).
Therefore,  at the horizon exit ($a\,H = k$), we find that
\beq
\l(\frac{V(\phi)}{\mpl^{4}}\r)_{a\,H\, =\, k}\,=\, g\,\l(\frac{k}{k_{_\ast}}\r)^{-\frac{2}{q - 1}},\label{eqn: V-at-H-exit}
\eeq
where $g$ is defined in Eq.~(\ref{eqn: g}).
Substituting Eq.~(\ref{eqn: V-at-H-exit}) in  Eq.~(\ref{eqn: Tensor PS model}) gives
\beq
\mathcal{P}_{_{T}}(k)\, =\,A_{_T} \l(\frac{k}{k_{_\ast}}\r)^{-\frac{2}{q - 1}},
\label{eqn: Tensor PS}
\eeq
where
\beq
A_{_T} = \l(\frac{2}{3\pi^{2}}\r)g.
\label{eqn: AT}
\eeq

From Eqs.~(\ref{eqn: scalar PS}) and  (\ref{eqn: Tensor PS}) we get
\beq
n_{_{T}}\,=\, n_{_{S}} - 1\, =\,  -\frac{2}{q-1}~,
\label{eqn: ns-our-model}
\eeq
where $n_{_{S}}$ and $n_{_{T}}$ are defined in Eqs.~(\ref{eqn: ns definition}) and (\ref{eqn: nt definition}), respectively. Note that the above expression for $n_{_{S}}$ and $n_{_{T}}$ is {\em independent} of the parameter $\alpha$ in the Lagrangian (\ref{eqn: Lagrangian}). In other words, one gets {\em the same result} for $n_{_{S}}$ and $n_{_{T}}$ for canonical ($\alpha =1$) and non-canonical ($\alpha \neq 1$) PLI\footnote{It is clear that in the slow roll limit (\ref{eqn: ns-our-model}) reduces to (\ref{eqn: ns-can}). The fact that  $n_{_{T}}\,=\, n_{_{S}} - 1\, \simeq\, -2/q$ in the slow roll regime, is a generic result for any power law inflation model based on the action (\ref{eqn: action}) but for which the parameter $\sigma$  defined in Eq.~(\ref{eqn: sigma}) vanishes, which corresponds to models with a constant speed of sound. This simply follows from the fact that for any inflationary model based on the action (\ref{eqn: action}), in the slow roll limit, one gets~\cite{Garriga-1999}: $n_{_{S}} - 1\, \simeq\, -4\varepsilon\, +\, 2\delta\, -\, \sigma$ and $n_{_{T}} \simeq -2\varepsilon$, where the slow roll parameters $\varepsilon$, $\delta$ and $\sigma$ are defined in Eqs.~(\ref{eqn: slow roll p 1}) and (\ref{eqn: sigma}), respectively. For the PLI model, since $\varepsilon \simeq 1/q$ and $\delta = \varepsilon$ one gets $n_{_{T}}\,=\, n_{_{S}} - 1\, \simeq\, -2/q$ whenever $\sigma$ vanishes, which is  the case for PLI models discussed in this paper. For PLI models with  $\sigma \neq 0$ but $|\sigma| << 1$ one gets $n_{_{T}}\,\neq\, n_{_{S}} - 1$, an example of this scenario is discussed in~\cite{Spalinski-2007}.}. It is also interesting to note that although we started with the expression (\ref{eqn: scalar PS model}) and (\ref{eqn: Tensor PS model}) which presumes slow roll, the result (\ref{eqn: ns-our-model}) is exact and one gets the same result even without imposing the slow roll condition~!

\begin{figure}[t]
\begin{center}
\scalebox{1.1}[1.2]{\includegraphics{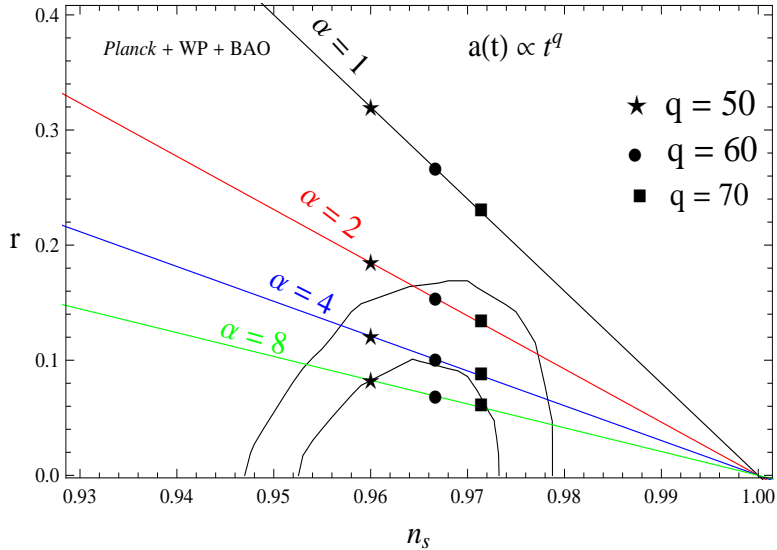}}
\caption{CMB constraints on power law inflation, $a(t) \propto t^q$, driven by the {\em inverse power law} potential (\ref{eqn: V-phi}). The spectral index $n_{_S}$ and the tensor-to-scalar ratio $r$ are shown for different values of the parameter $\alpha$ in (\ref{eqn: Lagrangian}). Here $\alpha\,=\, 1$ corresponds to canonical power law inflation~(\ref{eqn: can-Lagrangian}). The inner and outer contours correspond to $1\sigma$ and $2\sigma$ confidence limits obtained using the {\em Planck}~$+$~WP~$+$~BAO data. The non-canonical version ($\alpha > 1$) of PLI is in good agreement with CMB observations.}
\label{fig:ns-r}
\end{center}
\end{figure}

Moving on to the tensor-to-scalar ratio, we find from Eqs.~(\ref{eqn: T-to-S def}),
(\ref{eqn: scalar PS}), (\ref{eqn: AS-SR limit}), (\ref{eqn: Tensor PS}) and (\ref{eqn: AT}) that
\beq
r\, \simeq\, \f{16}{q\,\sqrt{2\,\alpha - 1}}~.
\label{eqn: T-to-S our model}
\eeq
Note that in obtaining the above result we have used Eq.~(\ref{eqn: AS-SR limit}) instead of Eq.~(\ref{eqn: AS}) as the difference between the two is insignificant in the slow roll limit $q\, >>\, 1$. Note that the tensor-to-scalar ratio {\em does depend} on the parameter $\alpha$ in the model~(\ref{eqn: Lagrangian}). In fact $r$ decreases as $\alpha$ is increased. It is for this reason that non-canonical inflation described by~(\ref{eqn: Lagrangian}) fares better than canonical PLI ~(\ref{eqn: can-Lagrangian}) which gives a larger value of $r$ than implied by \emph{Planck}. For example, when $q\, =\, 50$, Eq.~(\ref{eqn: ns-can}) and (\ref{eqn: r-can}) give $n_{_S} = 0.96$ and $r = 0.32$, respectively, for the canonical PLI model~(\ref{eqn: can-Lagrangian}), whereas $r = 0.096$ and $n_{_S} = 0.96$ for $q\, =\, 50$ and  $\alpha\, =\,6$ in (\ref{eqn: Lagrangian}).  Joint constraints on $n_{_S}$ and $r$ obtained from {\em Planck}~$+$~WP~$+$~BAO data are shown in the Fig.~\ref{fig:ns-r}. The non-canonical version of PLI is clearly in agreement with observations.

From (\ref{eqn: sound speed model}), (\ref{eqn: ns-our-model}) and (\ref{eqn: T-to-S our model}), one finds that in the slow roll regime ($q\,>>\,1$)
\beq
r\, = \, -8\,c_{_s}\,n_{_{T}} = 8\,c_{_s} (1-n_{_{s}})~.
\label{eq:consistency}
\eeq
This is the consistency relation for non-canonical PLI, and it distinguishes the class of models studied in this paper from canonical PLI for which $r\, = \, -8n_{_{T}}$.

\subsection{Non-gaussianity parameter $f_{_{\mathbf{NL}}}^{\mathrm{equil}}$}

Next we carry out a simple estimate of non-gaussianity in our model.
For a non-canonical model with the Lagrangian ${\cal L}(X,\phi)$, the non-gaussianity parameter $f_{_{\mathbf{NL}}}$ in the equilateral limit at leading order is given by~\cite{chen-2007}
\beq
f_{_{\mathbf{NL}}}^{\mathrm{equil}}\,=\, \frac{5}{81}\l(\frac{1}{c_{_s}^{2}}-1-\frac{2\lambda}{\Sigma}\r) - \frac{35}{108}\l(\frac{1}{c_{_s}^{2}}-1\r)~,\label{eqn: f-NL-def}
\eeq
where
\ber
\lambda\,&\equiv&\,X^{2}{\cal L},_{XX}\,+\,\frac{2}{3}X^{3}{\cal L},_{XXX}\nn\\
\Sigma\,&\equiv&\,X{\cal L},_{X}\,+\,2X^{2}{\cal L},_{XX}
\label{eqn: lambda-sigma-def}
\eer
Note that we use the WMAP sign convention in defining $f_{_{\mathbf{NL}}}^{\mathrm{equil}}$ which is opposite to that defined in \cite{chen-2007}.
One finds, from (\ref{eqn: Lagrangian}) and (\ref{eqn: lambda-sigma-def}) that
\beq
\frac{\lambda}{\Sigma} =  \f{\alpha-1}{3}~.
\label{eqn: l-sigma}
\eeq
Substituting this in (\ref{eqn: f-NL-def}) and using Eq.~(\ref{eqn: sound speed model}), we get
\beq
f_{_{\mathbf{NL}}}^{\mathrm{equil}}\,\simeq\,-0.57(\alpha-1)
\simeq -0.28\left(c_{_s}^{-2} - 1\right).
\label{eqn: f-NL-OPLI}
\eeq
For $\alpha = 6$, the above expression gives $f_{_{\mathbf{NL}}}^{\mathrm{equil}}\,\simeq -2.8$, which is in excellent agreement with the \emph{Planck} result $f_{_{\mathbf{NL}}}^{\mathrm{equil}}\, =\, -42\,\pm\,75$~\cite{Planck-NG}. It is easy to see that the \emph{Planck} bounds on $f_{_{\mathbf{NL}}}^{\mathrm{equil}}$ effectively translate into $1 \leq \alpha \leq 208$ for our model. However, as noted earlier, canonical PLI with $\alpha = 1$ is excluded on the basis of its values of $\lbrace n_{_S},~r\rbrace$. Indeed, the relationship between $n_{_S},~r$ and $c_{_s}$ in (\ref{eq:consistency}), together with the \emph{Planck} constraints: $f_{_{\mathbf{NL}}}^{\mathrm{equil}}\, =\, -42\,\pm\,75$, $n_{_{S}} \in [0.945-0.98]$ and $r\,<\,0.12$ at $95\%$ CL, allow one to infer that the sound speed
for PLI should lie in the range
\beq
0.05\,\lesssim\,c_{_s}\,\lesssim\,0.75\label{eqn: cs-range}
\eeq
which is valid when $c_{_s}$ is constant. Canonical PLI~(\ref{eqn: can-Lagrangian}) clearly does not satisfy the above criteria since $c_{_s} =1$ in this model.

However in our PLI model $c\,_{_s}\,=\,(2\alpha -1)^{-1/2}$, and the range~(\ref{eqn: cs-range}) is easily satisfied when $\alpha\,\gtrsim\,2$.
We therefore conclude that our new PLI model agrees very well with CMB observations for a wide range of values of the parameter $\alpha$.

\section{Exit from power law inflation}
\label{sec:exit}

A central drawback of power law inflation is that, since the universe accelerates forever, it cannot accommodate deceleration in the form of the radiative and matter dominated epochs which succeed inflation. A possible way out of this dilemma is to assume that the potential driving PLI approximates a more general potential which allows exit from inflation. Here we will demonstrate this possibility for the non-canonical PLI model described in the preceding section.

\begin{figure}[t]
\begin{center}
\scalebox{1.2}[1.2]{\includegraphics{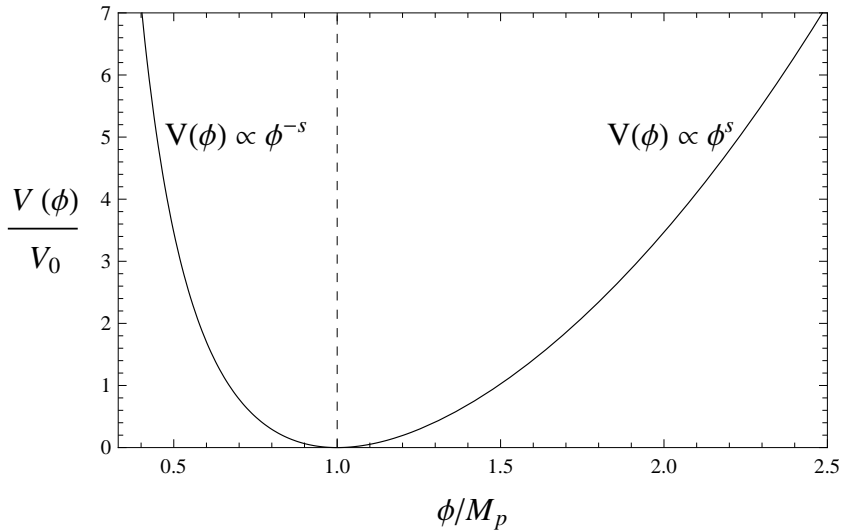}}
\caption{The potential~(\ref{eqn: exit V-phi}) is plotted as a function of $\phi$ for $\alpha = 6$. The vertical dashed line at $\phi\,=\,\mpl$ divides the potential into two branches. In the left branch ($\phi\,<\,\mpl$) the potential $V \propto \phi^{-s}$ leads to PLI, while in the right branch ($\phi\,>\,\mpl$) the potential becomes $V \propto \phi^{s}$ which corresponds to chaotic inflation. Inflation can successfully proceed from both branches of the potential.}
\label{fig:exit-pot}
\end{center}
\end{figure}

As noted earlier, the potential $V \propto \phi^{-s}$ drives PLI in the non-canonical setting~(\ref{eqn: Lagrangian}). The following slight change to this potential allows for PLI together with the possibility of a `graceful exit' from inflation:
\beq
V(\phi)\,=\,V_{_0}\l[\l(\frac{\phi}{\mpl}\r)^{-s/2}\;-\;\l(\frac{\phi}{\mpl}\r)^{s/2}\r]^{2};~~~s  = \frac{2\,\alpha}{\alpha - 1}~.
\label{eqn: exit V-phi}
\eeq
In Fig.~\ref{fig:exit-pot} this potential is plotted for $\alpha\,=\,6$. The potential has two branches: the left corresponds to $\phi\,<\,\mpl$ and gives rise to PLI with  $V \propto \phi^{-s}$, while the right branch ($\phi\,>\,\mpl$) yields the potential commonly associated with chaotic inflation, namely $V \propto \phi^{s}$.
The potential in (\ref{eqn: exit V-phi}) has a minimum at $\phi\,=\,\mpl$ where $V(\phi)\,=\,0$. Oscillations about this minimum allow for the universe to reheat \cite{Bassett-2005}. The viability of inflation from both branches of $V(\phi)$ is discussed below.

\subsection{Power law inflation from the left branch of $V(\phi)$}

The left branch is characterized by $\phi \leq \mpl$. At $\phi = \mpl$ the potential in (\ref{eqn: exit V-phi}) vanishes and from (\ref{eqn: rho-p-model}) one finds that the equation of state of the scalar field is
\beq
w_{_{\phi}} = \frac{p_{_{\phi}}}{\rho_{_{\phi}}} = \frac{1}{2\alpha-1} > 0\,,
\eeq
thus ensuring that the inflationary epoch has ended.

To determine {\em precisely when} inflation ends one needs to take a look at the slow
roll parameter $\varepsilon$. Substituting $V(\phi)$ from (\ref{eqn: exit V-phi}) into (\ref{eqn: potential SR1}) we get
\beq
\varepsilon_{_V} = \l[\frac{1}{\alpha}
\l(\frac{3\,M^{4}}{V_{_0}}\r)^{\alpha -1}\l(\frac{s\l(\phi_{_N}^{-s}-\phi_{_N}^{s}\r)}{\phi_{_N}\sqrt{2}}\r)^{2\alpha}
\l(\frac{1}{\phi_{_N}^{-s/2}-\phi_{_N}^{s/2}}\r)^{6\alpha-2}
\r]^{\frac{1}{2\alpha - 1}}~,
\label{eqn: potential SR exit}
\eeq
where $\phi_{_N}\,=\,\phi/\mpl$ and for PLI it is necessary that $V_{0}$ and $M$ be related \emph{via} Eq.~(\ref{eqn: V-0}). Eq. (\ref{eqn: potential SR exit}) implies $\varepsilon_{_V}\,\simeq\,1/q$ when $\phi_{_N}\,<<\,1$, and the slow roll regime ($q\,>>\,1$) is satisfied. This simply reflects the fact that $V \propto \phi^{-s}$ when $\phi_{_N}\,<<\,1$. As the field rolls further down its potential, the value of $\varepsilon_{_V}$ increases until $\varepsilon_{_V}\,=\,1$ which is reached when
\beq
\phi\,=\,\phi_{_e}\,=\,0.91\,\mpl~,
\label{eqn: phi end}
\eeq
for $\alpha\,=\,6$ and $q\,=\,50$. Therefore at this point inflation ends for this model.

Next, it is important to ascertain that the new potential~(\ref{eqn: exit V-phi})  does not alter the power law nature of the solution at roughly $60$ e-folds from the end of inflation. Otherwise, $n_{_S}$ and $r$ will differ from (\ref{eqn: ns-our-model}) and (\ref{eqn: T-to-S our model}), respectively.

The value of the field at $60$ e-folds from the end of inflation is determined by noting that
\beq
\frac{\d \phi}{\d N}\,=\,-\frac{\dot{\phi}}{H}~,
\label{eqn: d phi by dN-def}
\eeq
where $N$ is the number of e-folds counted from the end of inflation. In the slow roll regime
\beq
H^{2}\,\simeq\,\frac{V(\phi)}{3\,\mpl^{2}}~.
\label{eqn: H}
\eeq
Using Eq.~(2.28) of Ref.~\cite{our-JCAP-paper}, one finds from (\ref{eqn: d phi by dN-def}) and (\ref{eqn: H})
\beq
\frac{\d \phi_{_N}}{\d N\,}\,=\,\l[
\l(\frac{6\,M^{4}}{V_{_0}}\r)^{\alpha-1}
\l(\frac{s\,(\phi_{_N}^{-s}-\phi_{_N}^{s})\,}{\alpha\phi_{_N}}\r)
\l(\frac{1}{\phi_{_N}^{-s/2}-\phi_{_N}^{s/2}}\r)^{2\alpha}
\r]^{\frac{1}{2\alpha - 1}}~.
\label{eqn: eqn: d phi by dN}
\eeq
Integrating the above equation with the condition that at $N\,=\,0$, $\phi\,=\,\phi_{_e}$,  we find for $\alpha\,=\,6$ and $q\,=\,50$
\beq
\phi\,=\,0.24\,\mpl~~~~\mathrm{at}~~~N\,=\,60.
\label{eqn: phi initial}
\eeq
Substituting the above value of $\phi$ in Eq.~(\ref{eqn: potential SR exit}) gives
\beq
\varepsilon_{_V}\,=\,\frac{1.1}{q}~~~~~~\mathrm{at}~~~N\,=\,60\,,
\label{eqn: ev at end}
\eeq
which should be compared with $\varepsilon_{_V}\,=\,q^{-1}$ for the power law model (\ref{eqn: V-phi}).
For $N\,=\,70$, equations~(\ref{eqn: potential SR exit}) \& (\ref{eqn: eqn: d phi by dN}) give
$\varepsilon_{_V}\,=\,1.07/q$. We therefore conclude that the potential~(\ref{eqn: exit V-phi}) does not significantly alter the power law nature of the solution $a(t)\, \propto\, t^{q}$ between 60 \& 70 e-folds from the end of inflation. Consequently, $n_{_S}$ and $r$ described by (\ref{eqn: ns-our-model}) and (\ref{eqn: T-to-S our model}), remain valid for the potential~(\ref{eqn: exit V-phi}) when the field rolls down it along the left branch.

Next we proceed to determine the values of the parameters $V_{_0}$ and $M$ in our model using CMB normalization
\emph{viz.~}$P_{_{S}}(k_{\ast}) = 2.2\times10^{-9}$ at the pivot scale $k_{\ast} = 0.05\,\mathrm{Mpc}^{-1}$~\cite{Planck-inflation}. We assume that the pivot scale exits the horizon at $\sim 60$ e-folds from the end of inflation. Since $r = P_{_{T}}/P_{_{S}}$, one finds from (\ref{eqn: Tensor PS model}), (\ref{eqn: T-to-S our model}) and (\ref{eqn: exit V-phi}) that
\beq
\frac{V_{_0}}{\mpl^{4}}\,=\,\l(\frac{24\pi^{2}P_{_{S}}(k_{\ast})}{q\sqrt{2\alpha-1}}\r)
\l(\frac{1}{(\phi_{_i}/\mpl)^{-s/2}-(\phi_{_i}/\mpl)^{s/2}}\r)^{2},
\label{eqn: v0 by Mp4}
\eeq
where $\phi_{_i}$ is the field value $60$ e-folds from the end of inflation. For $\alpha\,=\,6$ and $q\,=\,50$, Eqs.~(\ref{eqn: phi initial}) and (\ref{eqn: v0 by Mp4}) give
\beq
V_{_0}\,=\,1.08\times 10^{-10}\,\mpl^{4}~.
\label{eqn: parameter v0}
\eeq
Substituting this value in (\ref{eqn: V-0}), we get
\beq
M\,=\,2.3\times 10^{-4}\,\mpl~.
\label{eqn: parameter M}
\eeq
From (\ref{eqn: exit V-phi}),  (\ref{eqn: phi initial}) and (\ref{eqn: parameter v0}), one gets $V(\phi) \simeq 3.14\times 10^{-9}\,\mpl^{4}$ and $H_{\ast} \simeq 3.2\times 10^{-5}\,\mpl$ for $\alpha\,=\,6$ and $q\,=\,50$.
Here $V$ and $H$ are evaluated at roughly $60$ e-folds from the end of inflation.

\subsection{Chaotic inflation from the right branch of $V(\phi)$}

The right branch of the potential~(\ref{eqn: exit V-phi}), corresponds to $\phi\,>\,\mpl$ and $V \propto \phi^{s}$ (see Fig.~\ref{fig:exit-pot}). The non-canonical version of chaotic inflation based on~(\ref{eqn: Lagrangian}) was described in detail in Ref.~\cite{our-JCAP-paper}. Since $s = 2\alpha/(\alpha-1)$, Eqs.~(3.13) and (3.23) of~\cite{our-JCAP-paper} imply
\ber
n_{_S}\,&=&\, 1 - \frac{2(4\alpha-3)}{4N(\alpha-1) + 2\alpha-1}\nn\\
r\,&=&\, \frac{16\sqrt{2\alpha-1}}{4N(\alpha-1)+2\alpha-1}~.
\label{eqn: ns-r chaotic}
\eer
For $\alpha = 6$, the above equation gives $n_{_S} = 0.965 $ and $r = 0.04$, respectively, at $60$ e-folds from the end of inflation  {\it i.e.~}at $N = 60$. These values are consistent with the \emph{Planck} results~\cite{Planck-inflation}.

The value of the model parameters $V_{_0}$ and $M$ was earlier fixed in the context of PLI which arises from the left branch of~(\ref{eqn: exit V-phi}). This procedure yielded $V_{_0}$ and $M$ given by Eqs.~(\ref{eqn: parameter v0}) and (\ref{eqn: parameter M}), respectively. It is interesting to verify whether these value of $V_{_0}$ and $M$  are consistent with inflation being realized from the right branch of~(\ref{eqn: exit V-phi}) {\it i.e.~}when $\phi\,>\,\mpl$ and $V \propto \phi^{s}$. For inflation with this potential, the value of $V_{_0}$  can be fixed using the CMB normalized value $P_{_{S}}(k_{\ast}) = 2.2\times10^{-9}$, for a given value of $M$, and vice versa. However, we assume that $V_{_0}$ and $M$ are related via  Eq.~(\ref{eqn: V-0}) and therefore either $V_{_0}$ or $M$ can be fixed using the CMB normalization.

Since $P_{_{T}} = r\,P_{_{S}}$, it turns out from Eq.~(\ref{eqn: Tensor PS model}) and (\ref{eqn: exit V-phi}) that
\beq
\frac{V_{_0}}{\mpl^{4}}=
\l(\frac{3\,\pi^{2}\,r\,P_{_{S}}(k_{\ast})}{2}\r){\l(\frac{1}{(\phi_{_i}/\mpl)^{-s/2}-(\phi_{_i}/\mpl)^{s/2}}\r)^{2}},
\label{eqn: V-0 chaotic}
\eeq
where $\phi_{_i}$ is the value of the field at $60$ e-folds from the end of inflation when inflation is realized from the right branch of potential~(\ref{eqn: exit V-phi}) and $r$ is given by Eq.~(\ref{eqn: ns-r chaotic}). For $\alpha = 6$ and $q = 50$, it turns out that $\phi_{_i}\,\simeq\,2\,\mpl$. On substituting this value of $\phi_{_i}$ in Eq.~(\ref{eqn: V-0 chaotic}), we get $V_{_0}\,\simeq 3.3 \times 10^{-10}\,\mpl^{4}$ and consequently Eq.~(\ref{eqn: V-0}) gives $M\,\simeq\,3 \times 10^{-4}\,\mpl $. These values are nearly close to those required by PLI in Eqs.~(\ref{eqn: parameter v0}) and (\ref{eqn: parameter M}), respectively. In fact when $\alpha = 4$ and $q = 60$, the CMB normalized value of $M$ as determined from both the branches of the potential~(\ref{eqn: exit V-phi}) is approximately the same and is given by $M \simeq 2\times 10^{-4}\mpl$.

We therefore conclude that both the branches of the potential~(\ref{eqn: exit V-phi}) give rise to viable inflationary models which are compatible with CMB results.

\section{Conclusions}\label{sec:conclusions}
Power law inflation in a spatially flat universe can be realized in a number of distinct ways:
\begin{itemize}
\item[(i)] By a minimally coupled canonical scalar field with an exponential potential~\cite{Lucchin-1985}
examined in Sec.~\ref{sec:canonical PLI}.
\item[(ii)] By k-inflation~\cite{Picon-1999,Garriga-1999}.
\item[(iii)] By a scalar field with a non-canonical kinetic term and with an inverse power law potential~(\ref{eqn: V-phi}), examined in Sec.~\ref{sec:new PLI}.
\end{itemize}
Scenario (i) appears to be in tension with recent CMB data due to the large value of $r$ in this model.
 By contrast scenario's (ii) and (iii) agree with the CMB data.

Our new model (iii) gives values for $n_{_S}$, $r$ and $f_{_{\mathbf{NL}}}^{\mathrm{equil}}$ which are in excellent agreement with CMB data for a wide range in parameter space.

Conventionally, power law inflation ($a \propto t^q$, $q > 1$) possesses a graceful exit problem, since cosmic acceleration never ends in this model. Sec.~\ref{sec:exit} provides a remedy to this problem by adding another branch to the inverse power law potential~(\ref{eqn: V-phi}). The new potential, given by (\ref{eqn: exit V-phi}), accommodates PLI along its left branch ($\phi\,<\,\mpl$) and chaotic inflation along its right branch ($\phi\,>\,\mpl$). Oscillations of the scalar field at the minimum value of $V(\phi)$ allow the universe to reheat. Thus the PLI model presented by us in this paper appears to be a viable model of inflation in all respects.

\section*{Acknowledgments}
We thank S.~Shankaranarayanan  and Debottam Nandi for  useful discussions
and acknowledge correspondence with Alexei Starobinsky and an anonymous referee which
improved the quality of the manuscript.
SU acknowledges the support of Max Planck-India Partner Group on Gravity and Cosmology.


\end{document}